\documentclass[preprint,12pt]{elsarticle}

\usepackage[latin1]{inputenc}

\usepackage{graphicx}
\usepackage{pdfpages}

\usepackage{subfigure}

\graphicspath{{Graphic/}}
\DeclareGraphicsExtensions{.pdf,.jpeg,.jpg,.png}

\usepackage{textcomp}

\usepackage[version=3]{mhchem}

\usepackage{array}
\usepackage{booktabs}

\usepackage{amssymb}
\usepackage{lineno}

\journal{Nucl. Instrum. Methods Phys. Res. Sect. A}

\begin{document}

\begin{frontmatter}



\title{A setup for soft proton irradiation of X-ray detectors for future astronomical space missions}


\author[IAAT]{Sebastian Diebold\corref{cor}}
\ead{diebold@astro.uni-tuebingen.de}
\author[ISDC]{Philipp Azzarello}
\author[INAF]{Ettore Del Monte}
\author[INAF]{Marco Feroci}
\author[IAAT]{Josef Jochum}
\author[IAAT]{Eckhard Kendziorra}
\author[IAAT]{Emanuele Perinati}
\author[INFN]{Alexandre Rachevski}
\author[IAAT]{Andrea Santangelo}
\author[IAAT]{Christoph Tenzer}
\author[INFN]{Andrea Vacchi}
\author[INFN]{Gianluigi Zampa}
\author[INFN]{Nicola Zampa}

\address[IAAT]{Kepler Center for Astro and Particle Physics, Universität Tübingen,\\Sand 1, 72076 Tübingen, Germany}
\address[ISDC]{ISDC Data Centre for Astrophysics, University of Geneva, Switzerland}
\address[INAF]{INAF/IASF Istituto di Astrofisica Spaziale e Fisica Cosmica, Rome, Italy}
\address[INFN]{INFN Istituto Nazionale di Fisica Nucleare, Sezione di Trieste, Italy}

\cortext[cor]{Corresponding author~~~\textit{Phone:} +49 7071 29 78609~~~\textit{Fax:} +49 7071 29 3458}

\begin{abstract}
Protons that are trapped in the Earth's magnetic field are one of the main threats to astronomical X-ray observatories. Soft protons, in the range from tens of keV up to a few MeV, impinging on silicon X-ray detectors can lead to a significant degradation of the detector performance. Especially in low earth orbits an enhancement of the soft proton flux has been found. A setup to irradiate detectors with soft protons has been constructed at the Van-de-Graaff accelerator of the Physikalisches Institut of the University of Tübingen. Key advantages are a high flux uniformity over a large area, to enable irradiations of large detectors, and a monitoring system for the applied fluence, the beam uniformity, and the spectrum, that allows testing of detector prototypes in early development phases, when readout electronics are not yet available.

Two irradiation campaigns have been performed so far with this setup. The irradiated detectors are silicon drift detectors, designated for the use on-board the \textit{LOFT} space mission.

This paper gives a description of the experimental setup and the associated monitoring system.
\end{abstract}


\end{frontmatter}

%\linenumbers

\section{Introduction}\label{sec:Intro}
Protons impinging on solid-state detectors on-board astronomical X-ray observatories pose a severe threat and can degrade the X-ray detection performance or even lead to a detector failure \cite{Nieminen_2000}. Soft protons in the energy range 0.1 - 1\,MeV are stopped near the detector surface, where they produce ionization, and, in particular, displacement damage. They are potentially more harmful than energetic particles, which are not stopped inside the detector, because the deposited energy is larger, and the radiation effects are concentrated in a small volume around the stopping point.

Ionization near the detector surface generates electron-hole-pairs in the field oxide (\ce{SiO_2}). Some of the holes drift towards the \ce{SiO_2}/\ce{Si} interface, where they create silicon dangling bonds. This leads to an increase of the surface component of the leakage current, whereas lattice defects increase the bulk leakage current. Consequences are a degradation of the energy resolution, and an enlarged charge transfer inefficiency (CTI) in front-illuminated CCDs (charge coupled devices). One well-studied example is the sudden increase in the CTI of the \textit{Chandra ACIS} front-illuminated CCDs during the first month of operation \cite{Chandra_ACIS}.

High proton fluxes occur during the passage through the Van Allen radiation belts and in near-equatorial low earth orbits, where protons are trapped in the geomagnetic field. X-ray detectors can easily be reached by soft protons through collimator holes and the openings of coded-masks. In X-ray observatories that use Wolter type optics, soft protons are funneled through the X-ray mirror shells and focused onto the detectors in the focal plane \cite{Nieminen_2000,Aschenbach_2007}.

To study the effect of soft proton radiation on a particular detector, and to estimate the performance degradation during the mission, irradiation setups at accelerator facilities are necessary. The usual procedure is the evaluation of the total ionizing dose (TID) and the non-ionizing energy loss (NIEL) that will occur during the mission, e.g. by performing a Monte Carlo simulation with the expected orbital fluence and spectral distribution. TID and NIEL are then reproduced in the laboratory with one or two proton energies and the representative fluence. An experimentally more complicated but physically straightforward approach is an irradiation with a reproduction of the orbit spectrum by composing it successively from several Gaussian spectra. 

Both procedures are possible with the setup presented in this paper, even though the composition of the orbit spectrum could be problematic for high fluences due to annealing of the damages during the irradiation. Annealing is a thermal rearrangement process of lattice atoms, and can restore the detector performance up to the initial values. Since radiation damages anneal much faster at room temperature than at the typical operating temperatures of X-ray detectors ($\sim -20\,\mathrm{^\circ C}$), this effect has to be considered.

In this contribution we discuss a setup for the irradiation of X-ray detectors with soft protons that has been constructed at the Van-de-Graaff accelerator facility of the Physikalisches Institut of the University of Tübingen. In particular, this setup is designed to reproduce the soft proton spectrum below 1\,MeV in low earth orbits (cf. Section~\ref{sec:ProtonFluxLEO}). Section~\ref{sect:AccSetup} briefly introduces the accelerator facility and gives a description of the experimental setup. Details of the monitoring system are presented in Section~\ref{sect:monitoring}. Up to now, the setup has been used in two irradiation campaigns to test the changes in the surface leakage current of a prototype of the silicon drift detector (SDD) for the \textit{LOFT} (Large Observatory For x-ray Timing) \cite{LOFT} space mission (cf. Section~\ref{sec:LOFT}).

\section{Soft proton flux in low earth orbit}\label{sec:ProtonFluxLEO}
\begin{figure}[t]
\centering
\includegraphics[width=.6\textwidth]{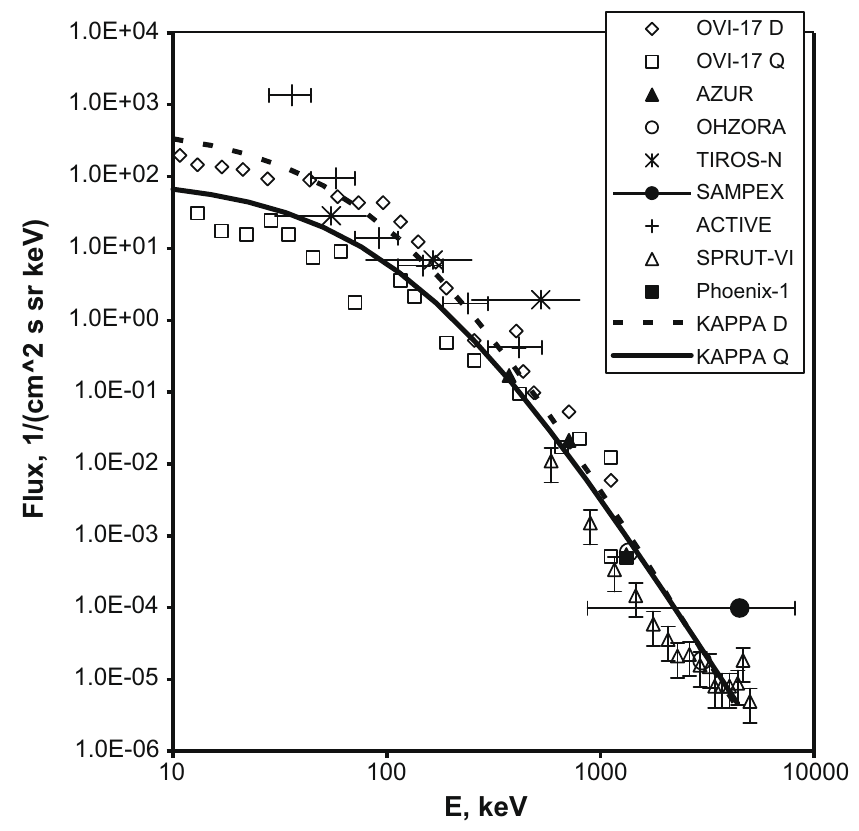}
\caption{The energy spectrum of protons in the near-equatorial region according to the data of several experiments \cite{Petrov_2009}. The measured spectrum was fitted using parameter sets for periods of quiet (solid line) and disturbed (dashed line) geomagnetic activity.}
\label{fig:Pspec_LEO_Petrov}
\end{figure}

A flux enhancement of soft protons has been found in the near-equatorial region at altitudes up to $\sim1300\,\mathrm{km}$. These soft protons could evolve via a double charge-exchange mechanism from the interaction of energetic protons with neutral atoms in the upper atmosphere (thermo- or exosphere) \cite{Moritz_1972}. A number of consistent measurements of this phenomenon show proton energies from $\sim10\,\mathrm{keV}$ to several MeV (cf. Figure~\ref{fig:Pspec_LEO_Petrov}). A combined analysis of these measurements is described in \cite{Petrov_2009}. It is proposed that two parameter sets, for quiet and disturbed conditions of the geomagnetic field, fit the experimental data best. The reported flux has been used to determine the dimensions of the irradiation setup, so that the fluence for a typical five year space mission can be applied in a reasonable time between some minutes and a few hours.

% --------------------------------------------------------------------------------------

\section{Irradiation setup}\label{sect:AccSetup}

\subsection{Accelerator facility}
% Rosenau accelerator
The irradiation setup has been constructed at the accelerator facility of the Physikalisches Institut of the University of Tübingen. The accelerator is a single ended $3\,\mathrm{MV}$ Van de Graaff (HVEC Model KN), that can provide light ion beams with energies ranging from $\sim700\,\mathrm{keV}$ to currently $2.4\,\mathrm{MeV}$. The reduced upper limit, compared to the design voltage, is due to an insufficient portion of sulfur hexafluoride (\ce{SF6}) in the protective gas, resulting in a lowered dielectric strength. It is planned for the near future to increase the portion of \ce{SF6}, so that $3\,\mathrm{MeV}$ can be reached. The lower limit of the beam current is of the order of a microampere.

% Voltage measurement and calibration
The terminal voltage is measured with a generating voltmeter \cite{IonBeamGVM}. The calibration of this voltmeter has been confirmed by measuring the $\gamma$ rate of the reaction \ce{^{27}Al}(p,$\gamma$)\ce{^{28}Si}, which has a sharp resonance in the cross section at 992\,keV, with a NaI(Tl) detector \cite{AccCalib}. 

% Ion source 
Four different gases are available for the ion source: hydrogen, deuterium, helium, and \ce{^{13}C}-enriched \ce{CO_2}. The radio frequency plasma source ionizes the selected gas only to the charge state 1+. If hydrogen is ionized, an equilibrium between protons and \ce{H2+} arises. An analyzing magnet with a deflection angle of $95^{\circ}$ allows to select the desired ions. By splitting \ce{H2+} with a thin foil or on the target, protons with half the nominal energy are obtained \cite{MolHydrogenBeams} (the same applies for deuterium). As \ce{H2+} and deuterons have the same charge-to-mass ratio, a momentum separation in the analyzing magnet is not possible, and therefore, the molecular hydrogen beam is contaminated with $\sim1\%$ deuterons. For this reason, the use of a proton beam is favored and metal energy degrader foils are used to lower the proton energy appropriately.

% Vacuum
Several combinations of rotary vane pumps and turbomolecular pumps are distributed along the beam line to reach a pressure in the $10^{-6}\,\mathrm{mbar}$ regime. This low pressure is especially necessary to avoid coating of the degrader foils with carbon during long duration irradiations, as the residual gas contains a certain amount of hydrocarbons. These molecules are cracked by the proton beam and free carbon is produced, that is then deposited on all surfaces, including the metal foils. Although the growth rate is small, the carbon layer alters the transmission properties of the foil and should be avoided. A beam stopper and a vacuum shutter in front of the foils give access to the experimental setup without venting the whole beam line, which would require a shut down of the accelerator. It takes about one hour to re-establish the vacuum in the experimental setup.

\subsection{Experimental setup}
% Beamline setup, alignment
The facility possesses six beam lines; number 3 is currently used for the irradiation setup. The beam line, including the position and opening of the slits and the position of the detector chamber, has been aligned with a theodolite. The beam can be bent and shifted in parallel with various dipole magnets and focused with two double quadrupoles.

\begin{figure*}[tb]
\centering
\includegraphics[width=.9\textwidth]{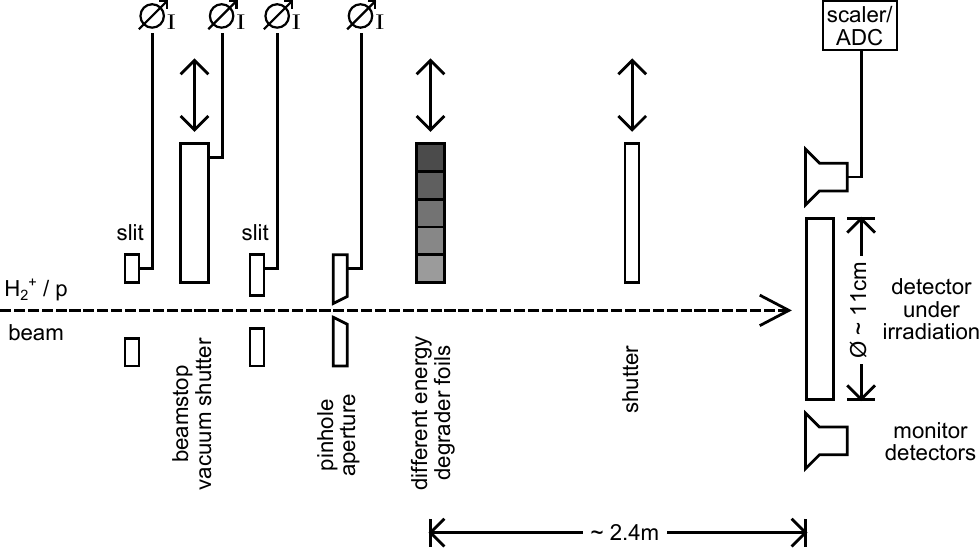}
\caption{Schematic of the irradiation setup.}
\label{fig:Setup_scheme}
\end{figure*}

\begin{figure*}[tb]
\centering
\includegraphics[width=\textwidth]{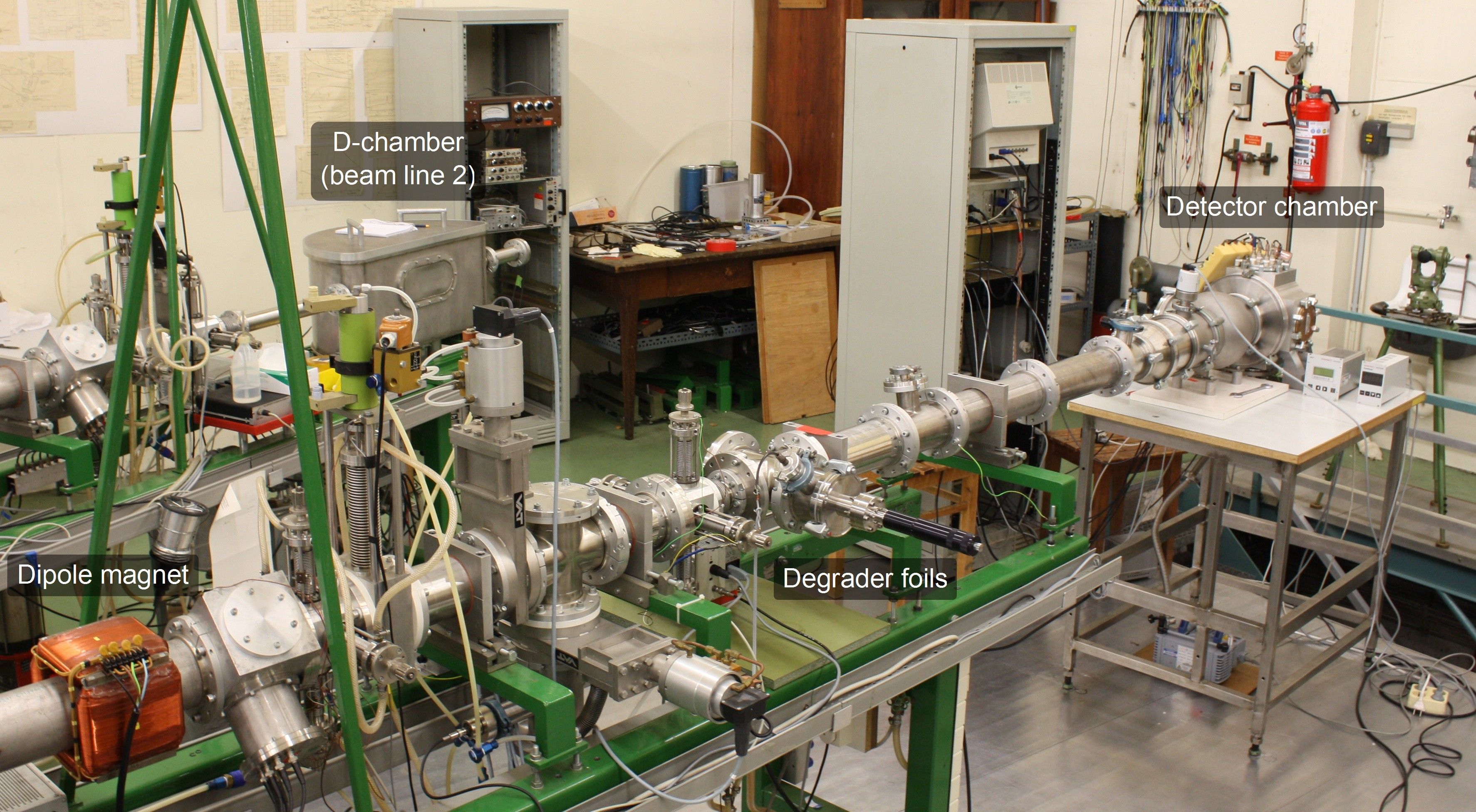}
\caption{Picture of the irradiation setup. The beam is coming in from the left where the last dipole magnet is visible. The black bar in the center is the linear manipulator on which the holder for the degrader foils is mounted. The detector chamber is located 2.4\,m downstream of the degrader foils, on the far right of the picture. The D-shaped chamber in the background (beam line 2) has been used for calibrating the monitor detectors and for measuring foil thicknesses.}
\label{fig:setup_picture}
\end{figure*}

% Slits
A schematic of the irradiation setup is presented in Figure~\ref{fig:Setup_scheme}, and the actual implementation at the accelerator facility is shown in Figure~\ref{fig:setup_picture}. First of all, the incoming proton beam passes through two slits. These slits are rectangular apertures, consisting of four isolated parts. From each of these parts a current signal can be tapped, which is used to align the beam position by comparing the signals from opposite slit parts. A straight beam in the center of the beam line is obtained if all opposite currents of the two slits are equal. The openings have been adjusted to $\sim4 \times 4\,\mathrm{mm^2}$ (first slit) and $\sim3 \times 3\,\mathrm{mm^2}$ (second slit).

% Pinhole aperture
A pinhole aperture with a circular opening assures a well defined beam spot on the degrader foil, and further reduces the beam current. Tests have shown that for low fluxes ($10^3$ - $5\cdot10^5\,\mathrm{cm^{-2}s^{-1}}$) an aperture with $0.1\,\mathrm{mm}$ is suitable. If fluxes of more than $\sim5\cdot10^5\,\mathrm{cm^{-2}s^{-1}}$ are required, the dia\-me\-ter must be enlarged accordingly. The aperture is made of copper, isolated from the beam line, and electrically connected to a vacuum feed-through, so that the current on the aperture can be monitored. The pinhole has a $45^{\circ}$-chamfer on the downstream (detector chamber) side to minimize the probability of small angle scattering inside the aperture, which would introduce a beam component with lower energy. 

\subsection{Degrader foils}
Thin metal foils with some micrometer thickness degrade and broaden the beam energy and widen the beam spatially. The energetic and spatial broadening is due to straggling. Four different foils can be fixed on a holder. The holder itself is mounted on a linear manipulator to allow a quick change of the foil without breaking the vacuum, e.g. for an irradiation with different energies, or to compose a spectrum similar to the in-orbit spectrum.

\begin{figure*}[tb]
\centering
\subfigure[simulated energy spectrum]{\includegraphics[width=.49\textwidth]{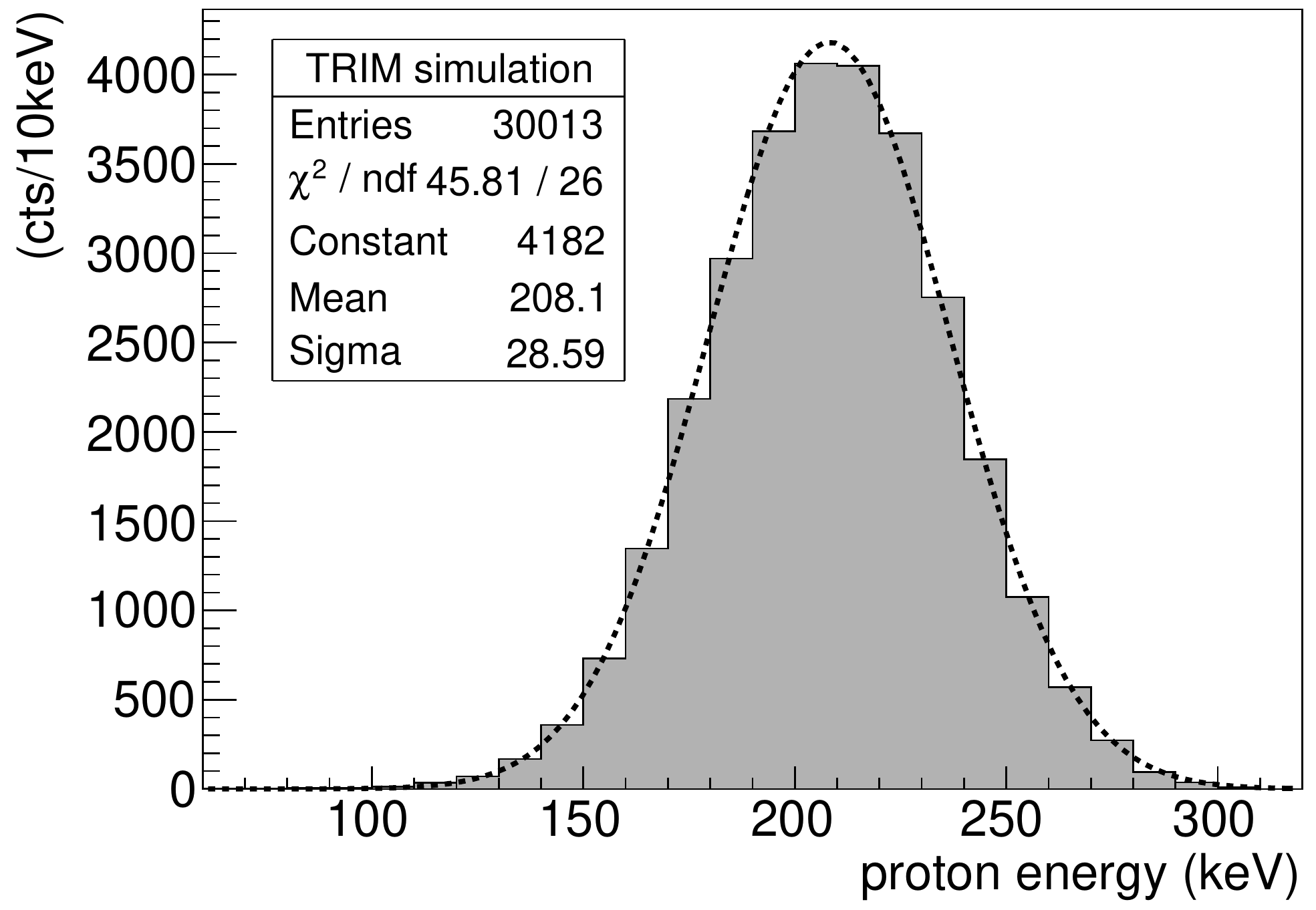}\label{subfig:TRIMspec}}
\subfigure[simulated angular distribution]{\includegraphics[width=.49\textwidth]{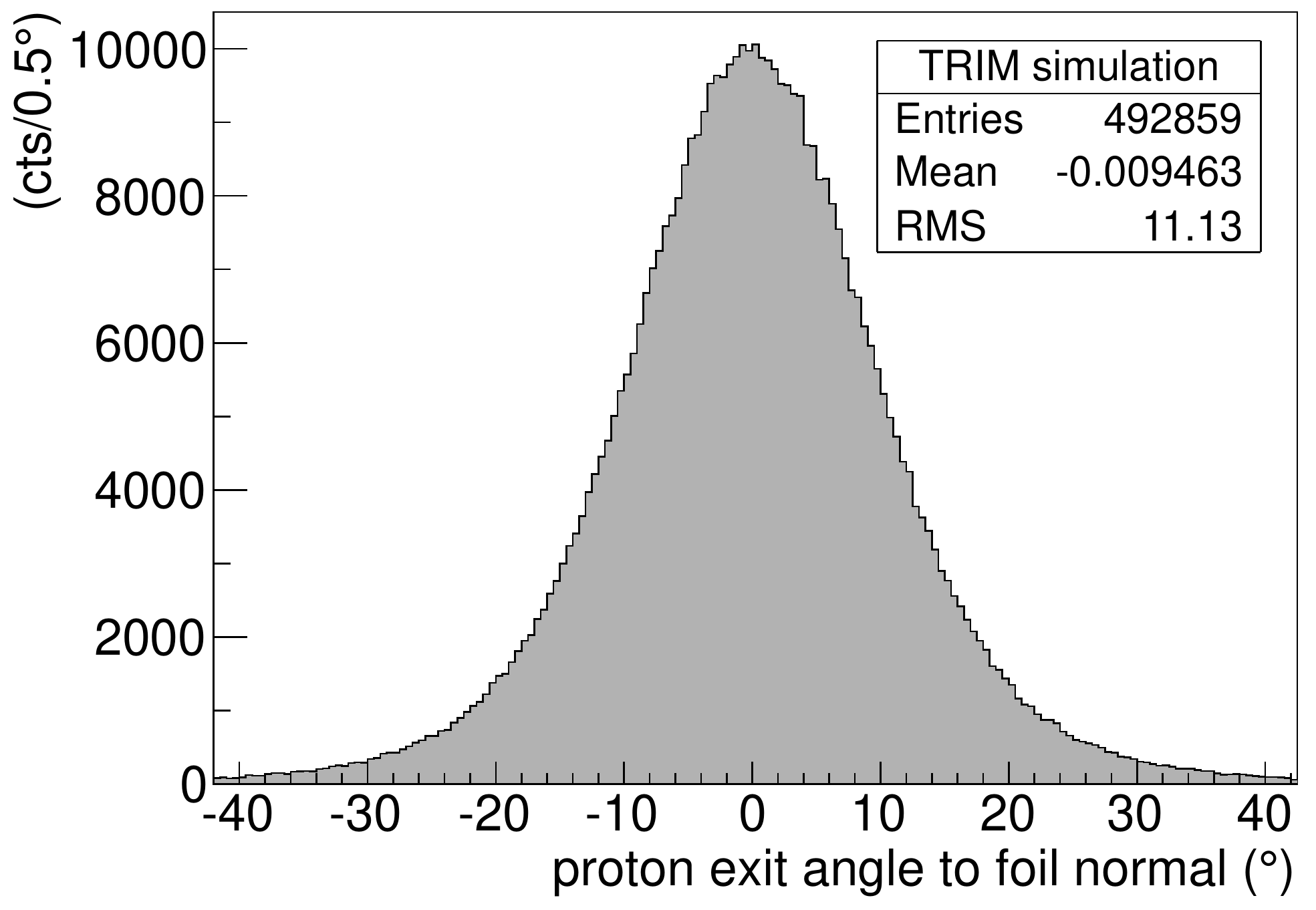}}
\caption{Results from a TRIM simulation with 1010\,keV beam energy and a 6\,\textmu m Cu foil. An angular cut of $3^\circ$ has been applied to the energy spectrum. The dotted line is a Gaussian fit to the spectrum.}
\label{fig:SimSpec}
\end{figure*}

For the selection of foil material, foil thickness, and beam energy to obtain a certain proton spectrum and spatial distribution, simulations with the TRIM\footnote{Transport and Range of Ions in Matter} Monte Carlo code (part of the SRIM\footnote{Stopping and Range of Ions in Matter} package \cite{SRIM}) have been carried out. The TRIM output lists energy and exit angle with respect to the foil normal for each transmitted proton. A software toolchain for the analysis of the simulations has been implemented. Since only protons with an exit angle of $\le 1.8^\circ$ can directly reach the detector chamber, an angular cut is applied to the simulation data. Without this cut the spectra are shifted to lower energies because of the contribution of protons which have been undergoing a larger energy transfer. The distribution of exit angles is used to estimate the flux uniformity in the detector chamber. As a consistency check, some of the foil - beam energy combinations have been simulated also by means of the Geant4 framework \cite{Geant4}, using the Livermore low energy electromagnetic physics list. The deviation in the mean energy is for most parameter combinations within 10\% of the TRIM output, and therefore within the expected accuracy. An interesting finding is that Geant4 systematically produces a lower mean energy and a broader spectrum than TRIM, but both values lie within systematic uncertainties of the detector calibration and foil thickness. A selection of these simulation results is presented in Table~\ref{tab:simout}. A typical spectrum and angular distribution simulated with TRIM is shown in Figure~\ref{fig:SimSpec}.

\begin{table}[tb]
\caption{Selection of typical simulation results for energy degrader foils, obtained with TRIM and Geant4. The values for mean energy and FWHM are derived from fitting the simulated spectra with Gaussian distributions.}
\label{tab:simout}
\centering
\begin{tabular}[c]{cccccc}\\
\toprule
		$E_{\mathrm{beam}}$ & Foil & $E_{\mathrm{mean,TRIM}}$ & $E_{\mathrm{FWHM,TRIM}}$ & $E_{\mathrm{mean,Geant4}}$ & $E_{\mathrm{FWHM,Geant4}}$\\
		(keV) &  & (keV) & (keV) & (keV) & (keV)\\ \midrule
		1010 & 6\,\textmu m Cu & 208 & 71 & 181 & 89\\
		2300 & 18\,\textmu m Cu & 820 & 115 & 781 & 130\\
		1010 & 12\,\textmu m Al & 308 & 54 & 290 & 68\\
		1010 & 14\,\textmu m Al & 120 & 75 & 116 & 111\\
\bottomrule
\end{tabular}
\end{table}

If the mean energy of the transmitted protons is larger than $\sim100\,\mathrm{keV}$, the spectra are very close to Gaussian distributions, independent of the foil parameters and the beam energy. For a given foil the spread in energy and exit angle is maximal if the beam energy is just above the transmission threshold, and decreases steadily with increasing beam energy. Straggling effects are larger in high-Z materials than in low-Z materials. This means that spectra with similar spectral width and angular distribution (flux uniformity) but different mean energies can be obtained by using low-Z materials for low proton energies and high-Z materials for high proton energies.

\begin{figure*}[tb]
\centering
\includegraphics[width=.55\textwidth]{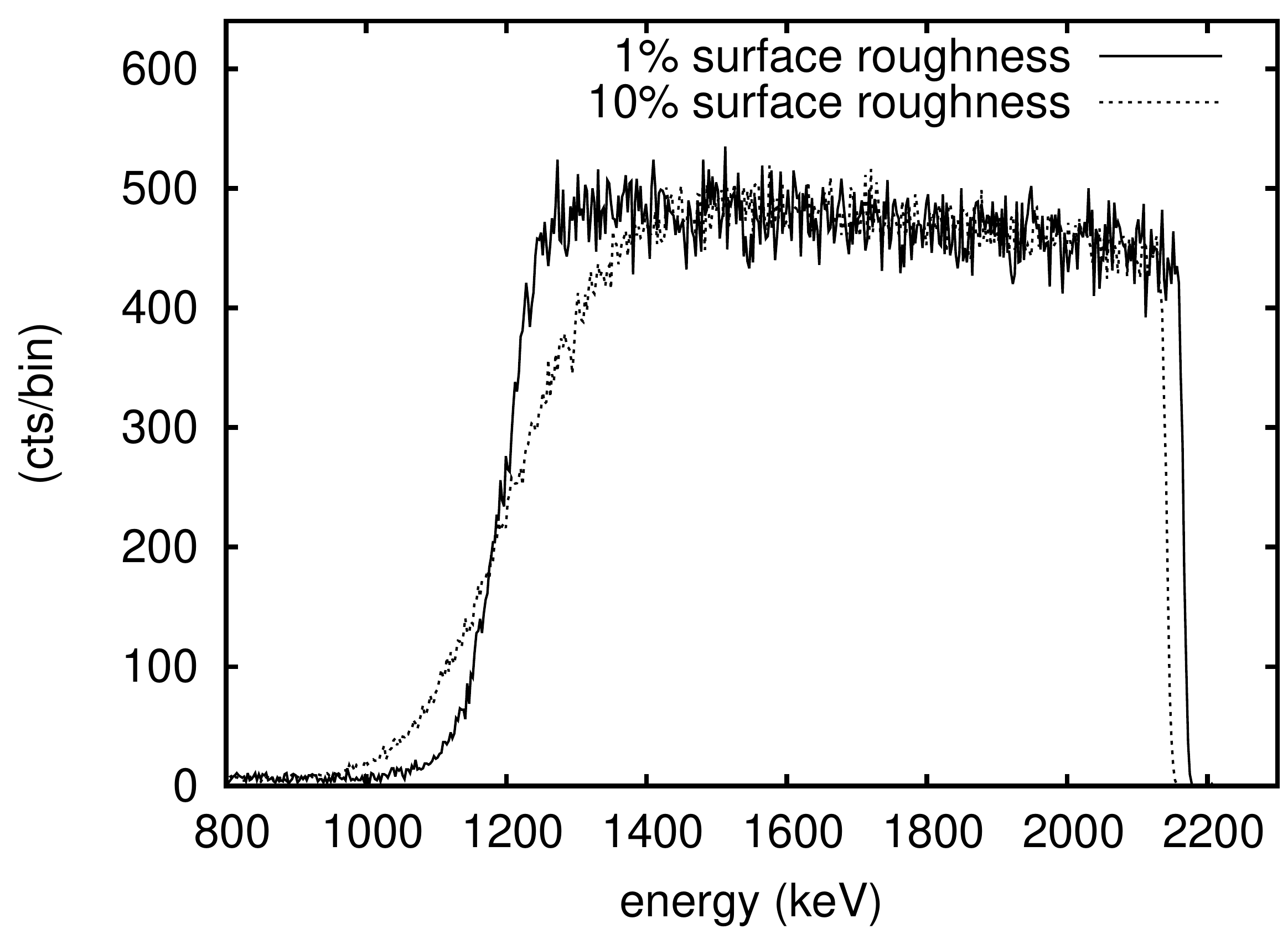}
\caption{RBS spectra of Cu foils with a nominal thickness of 6\,\textmu m from different ma\-nu\-facturers. Beam energy 2.3\,MeV, scattering angle $150^\circ$. The size of the RBS box is a measure for the foil thickness. The slope of the low energy edge is dependent on the surface roughness. A steeper edge indicates lower roughness.}
\label{fig:sur_rough}
\end{figure*}

The thicknesses of foils from two different manufacturers have been measured by means of Rutherford Backscattering Spectrometry (RBS). A beam of monoenergetic protons is targeted on the foil and two surface barrier detectors record the energy of backscattered protons. These detectors are placed at angles of $150^\circ$ and $165^\circ$ with respect to the incoming beam. If the beam energy is sufficient, e.g. $\gtrsim1800\,\mathrm{keV}$ for 6\,\textmu m \ce{Cu}, the spectrum shows a box-shaped feature, the so-called RBS box (cf. Figure~\ref{fig:sur_rough}). The width of this RBS box can be converted to the foil thickness. The significance of the method is dependent on the detector calibration and the stopping power data for the foil material. The typical error is 2-5\%. The SIMNRA\footnote{SIMulation code for Nuclear Reaction Analysis} code has been used to analyze the RBS spectra \cite{SIMNRA}.

Not only the thickness but also the surface roughness can be determined with RBS by examining the slope of the low energy edge of the RBS box. This edge is always less steep than the high energy edge because of straggling, where a steeper slope indicates a smoother surface. An example of the surface roughness deviation between the two foil manufacturers is represented in the backscattering spectra in Figure~\ref{fig:sur_rough}. A foil with higher surface roughness produces a broader spectrum for transmitted protons while maintaining the same mean energy.

\subsection{Measured proton spectrum}
\begin{figure*}[tb]
\centering
\subfigure[without anti-scatter apertures]{\includegraphics[width=.49\textwidth]{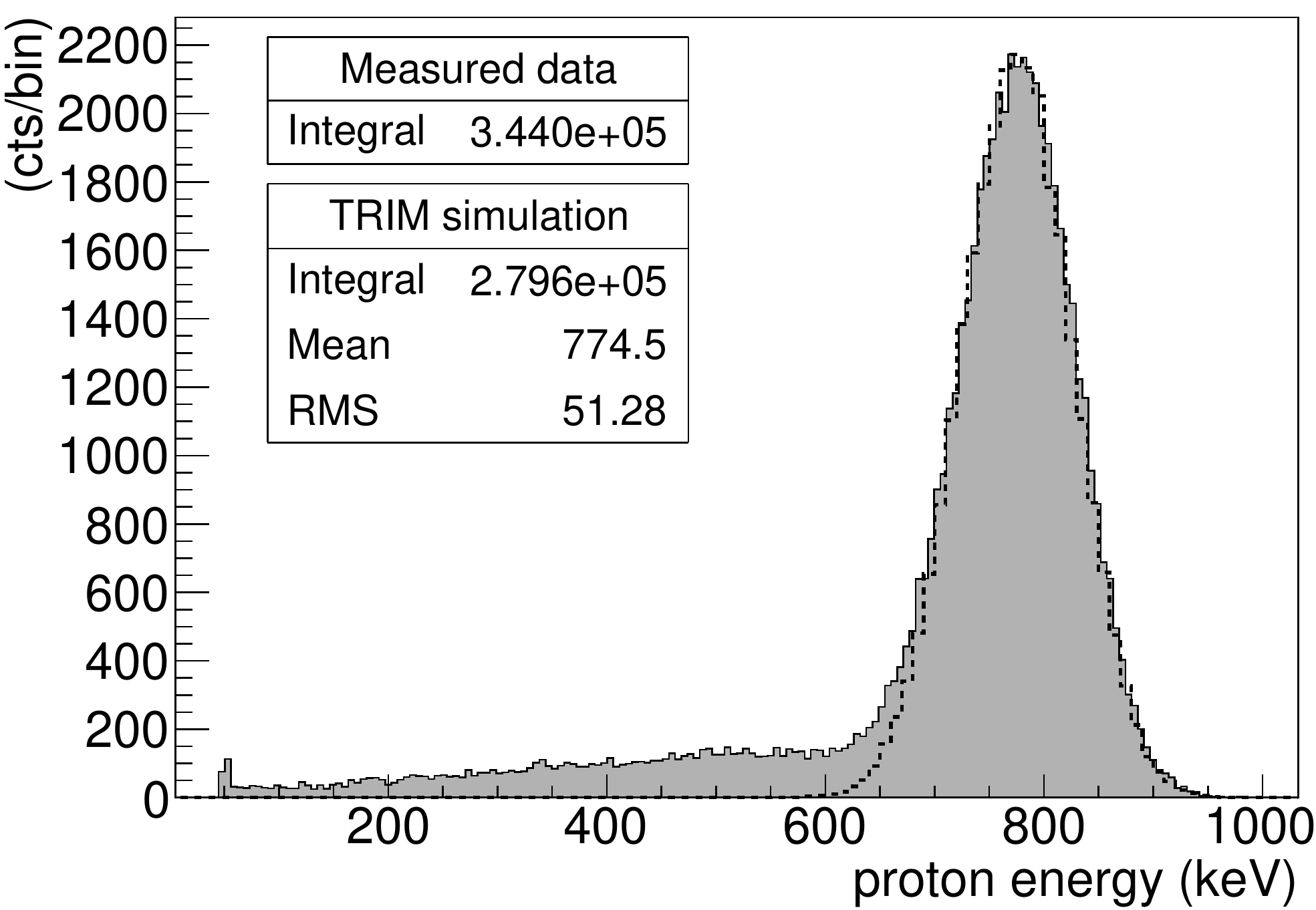}\label{fig:wo_asca}}
\subfigure[with anti-scatter apertures]{\includegraphics[width=.49\textwidth]{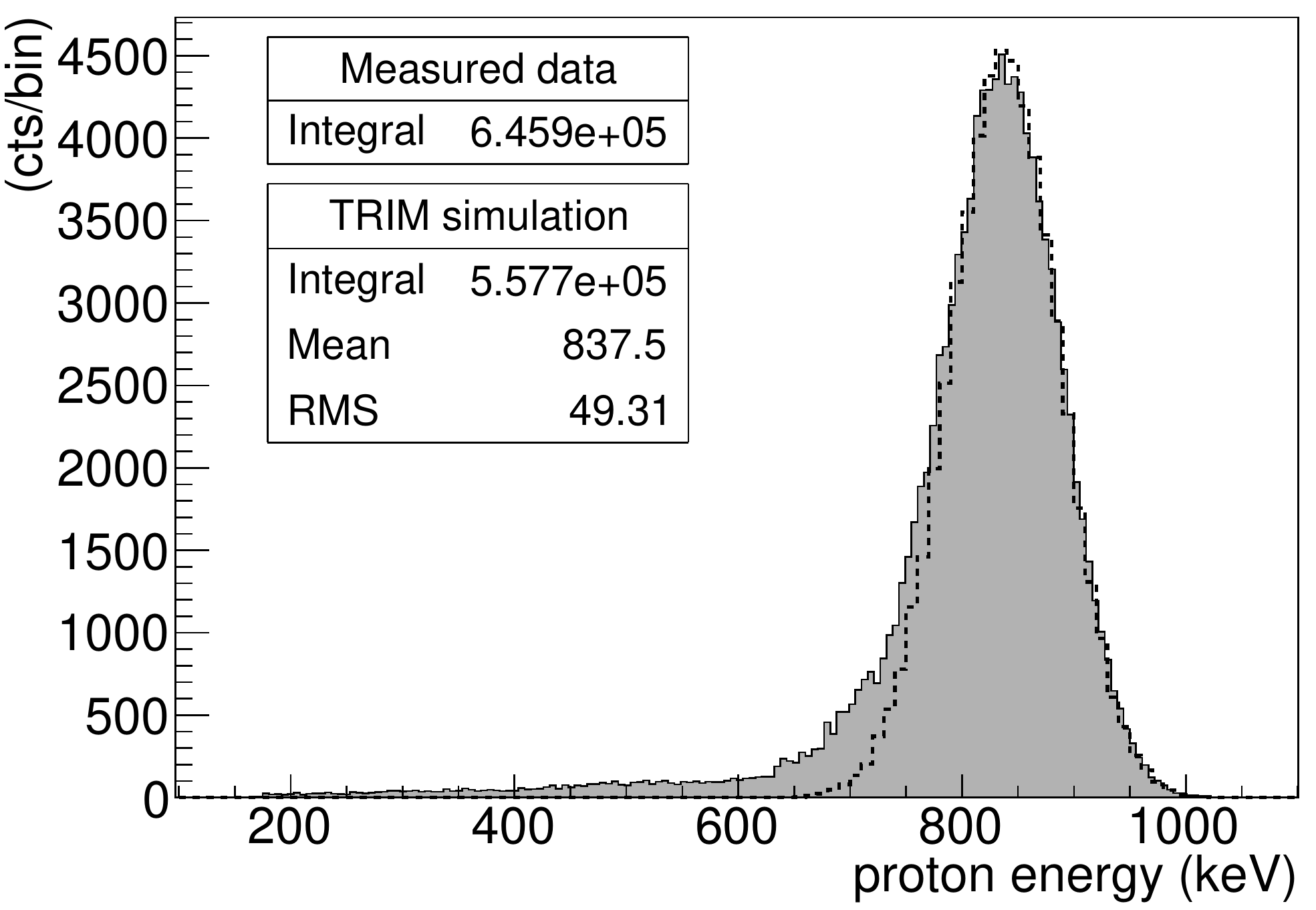}\label{fig:w_asca}}
\caption{Comparison of the spectrum with and without anti-scatter apertures. Both are measured with an 18\,\textmu m Cu foil. The beam energy is about 2.3\,MeV, but slightly different for the two measurements, resulting in a shift of the mean energy. The dotted lines represent TRIM simulations with appropriate energies. In the spectrum with anti-scatter apertures the main peak shows an exponential decay at the low energy edge. This is due to forward scattering in the unsensitive top layers of the silicon surface barrier detectors. It is present in the spectrum without anti-scatter apertures as well but barely visible.}
\end{figure*}

% Anti-scatter apertures
The beam line between the foils and the detector chamber has a diameter of 10\,cm. Since this setup is intended for the irradiation of detectors larger than that, the last 1\,m of the beam line has been enlarged to 15\,cm diameter. Test measurements have shown that the Gaussian shape of the proton spectra, obtained from simulations, is well reproduced except for a low energy tail (cf. Figure~\ref{fig:wo_asca}). This tail arises because a fraction of protons with exit angles larger than $1.8^\circ$ are scattered from the inner walls of the beam line onto the irradiation plane. They contribute with a fraction of up to 20\% to the spectrum measured in the detector chamber. As a consequence, two 2\,mm aluminum apertures are inserted between the foils and the detector chamber, one with an opening of 3.6\,cm at a distance of 59\,cm from the foils and a second one with 8.4\,cm diameter at 137\,cm. These apertures define a cone with an opening angle slightly smaller than the $3.6^\circ$ defined by the beam line geometry. The fraction of protons with lower energies is drastically reduced to about 4\% (cf. Figure~\ref{fig:w_asca}). Geant4 simulations show that these 4\% arise mostly from the pinhole and the detector aperture (cf. Section~\ref{sect:monitoring}).

\section{Beam monitoring}\label{sect:monitoring}
\begin{figure*}[tb]
\centering
\includegraphics[width=.5\textwidth]{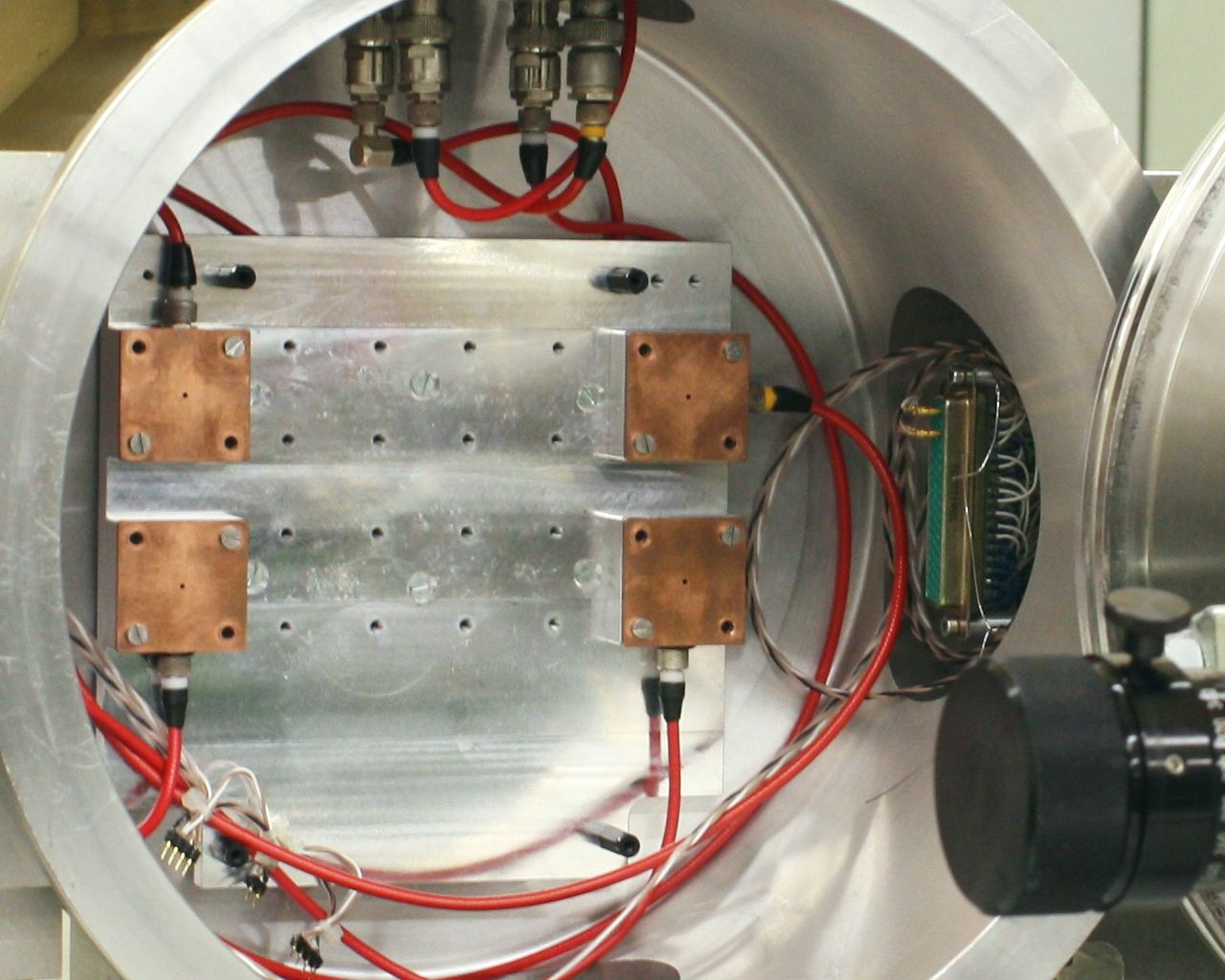}
\caption{Picture of the detector chamber (25\,cm inside diameter) with four SSB detectors at the peripheral monitor positions. Various mounting possibilities around the center allow a measurement of the flux uniformity and a determination of the offset to the peripheral positions. Copper apertures define the effective areas of the SSBs.}
\label{fig:PIC_monitors}
\end{figure*}

% Mounting of test detector and flux monitoring
Up to four silicon surface barrier (SSB) detectors can be used to monitor the proton flux and spectrum, and for the measurement of the beam uniformity. Copper apertures define the effective area of each detector. An aluminum board on the backside of the detector chamber provides various mounting possibilities: either peripheral to monitor the fluence during an irradiation, or at the position of the detector (before an irradiation) to determine the offset from the center and to obtain a beam uniformity map (cf. Figure~\ref{fig:PIC_monitors}). In order to correct for dead-time effects of the data acquisition, a pulser signal is added to the signals of the monitor detectors and, simultaneously, given on a separate scaler. The acquisition efficiency is obtained by comparing the number of counts from the pulser in the spectra of the monitor detectors to the scaler value.

% Aluminum shield
Since a perfect reproduction of the beam properties is very challenging, it is crucial to measure the actual proton flux just before an irradiation. If necessary, the beam intensity can be adjusted by changing the extraction voltage of the ion source or by (de-)focusing the beam. In order to shield the test detector during this measurement, a 2\,mm thick aluminum plate can be placed in front with a rotation manipulator, leaving just the monitor detectors exposed to the beam.

\subsection{Flux monitoring and homogeneity}
% Determination of effective area
The most critical parameter for a precise flux measurement is the determination of the openings of the copper apertures in front of the monitor detectors, since they define the effective detector areas. The diameters have been measured by using an x-y-table with \textmu m-position accuracy and an attached microscope with cross hairs. This method determines the effective area to about 2\%, which poses the limit on the precision of an absolute flux measurement.

% Homogeneity map
For the measurement of the beam uniformity the 2\% precision would only yield an upper limit, since for most useful combinations of foil parameters and beam energy the predicted non-uniformity from TRIM simulations is also of the order of a few percent. Significant measurements become possible by comparing the effective areas of the different detectors. This has been achieved by measuring the rate on two detectors, which are mounted symmetrical to the beam line center, then exchanging the positions and measuring the rate again. The error of this method has been determined to be about 0.3\% by repeating some of the measurements.

\begin{figure*}[tb]
\centering
\includegraphics[width=.7\textwidth]{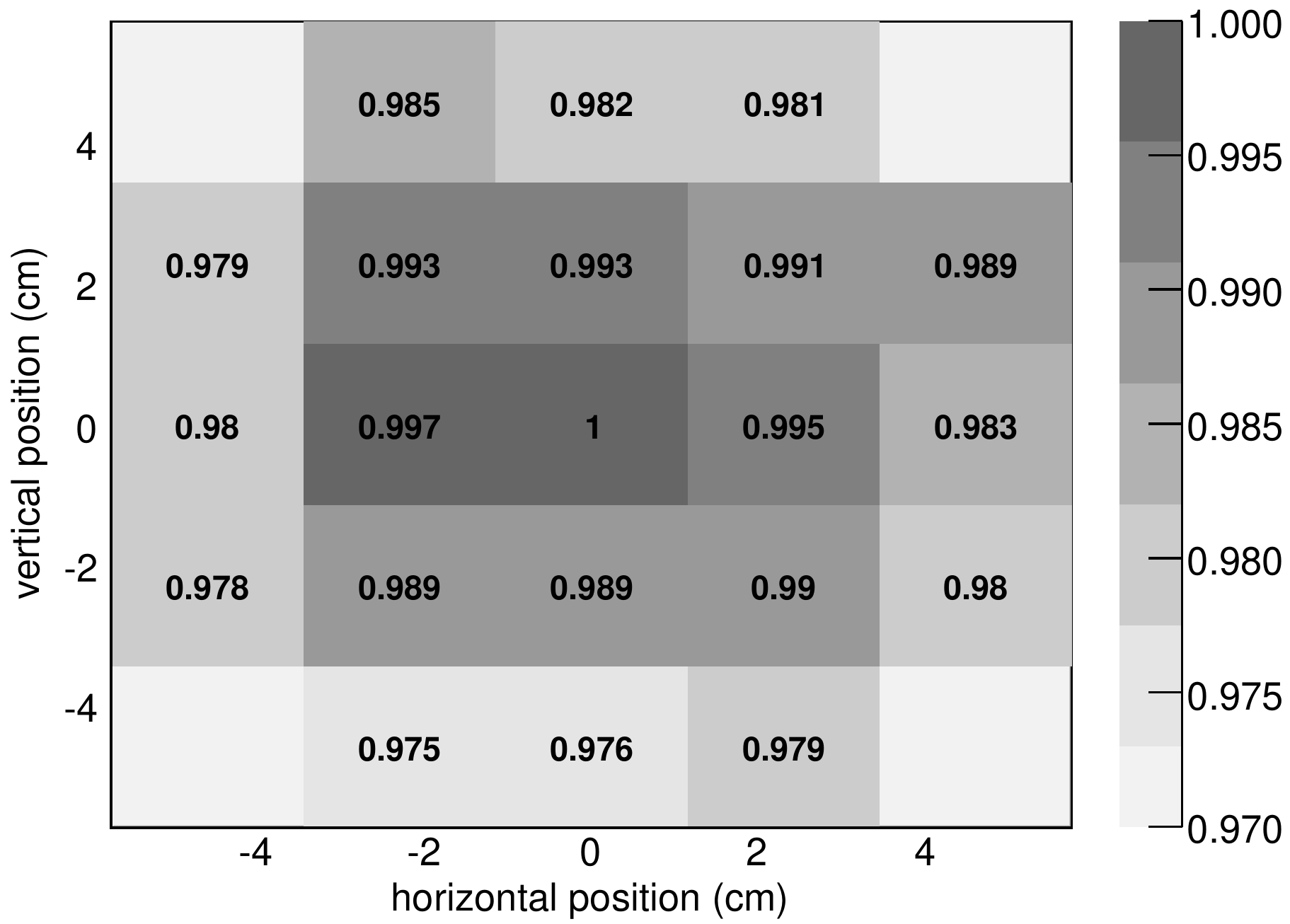}
\caption{Measured flux homogeneity map for 2.3\,MeV beam energy and an 18\,\textmu m Cu foil. The given values are normalized to the center of the beam.}
\label{fig:homogeneity}
\end{figure*}

During a homogeneity measurement the positions of two detectors are altered while two detectors remain at fixed positions. The rates of the moved detectors are then normalized to the mean rate of the fixed ones, to address the problem of fluctuations in the beam current. As an example, a homogeneity map for a beam energy of 2.3\,MeV and an 18\,\textmu m copper foil is shown in Figure~\ref{fig:homogeneity}. The flux deviation over an area of $8\times8\,\mathrm{cm^2}$ is less than 3\%. Such a high uniformity over this large area is a real advantage of the presented setup.

\subsection{Spectrum monitoring}
In order to reduce the dependence on simulations for the applied proton spectra the monitor detectors have been energy calibrated. Since SSBs have an insensitive layer on the front for the electric contact, a calibration with an alpha source, e.g. 5.5\,MeV alphas from \ce{^{241}Am}, could not be downscaled to the energy range of interest. The lowest proton energy available from the accelerator is 350\,keV if a molecular hydrogen beam is used, but the beam current is by far too large to point it directly on a SSB.

\begin{figure*}[tb]
\centering
\includegraphics[width=.6\textwidth]{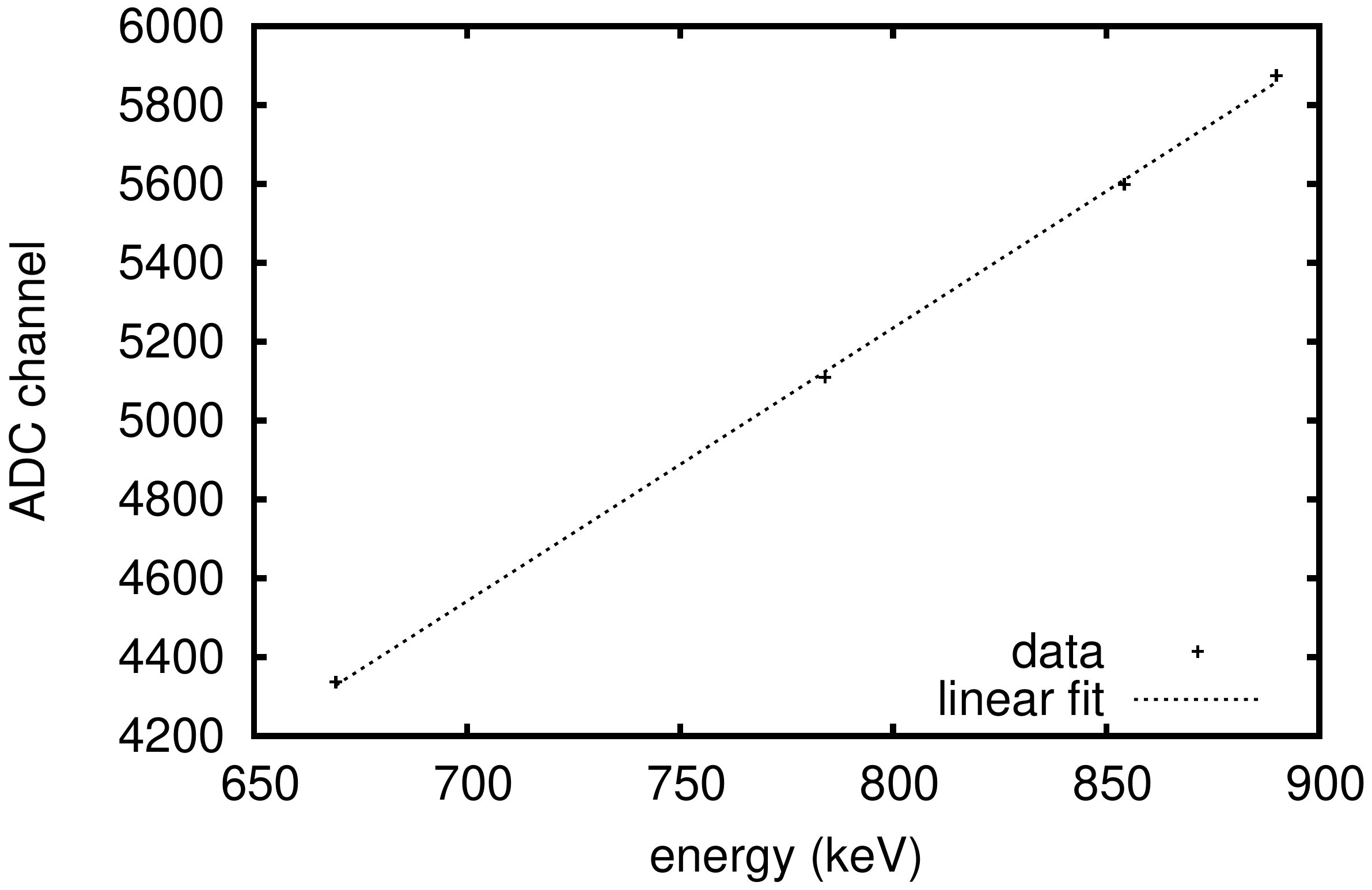}
\caption{Calibration of one of the monitor detectors with $165^\circ$ backscattered protons from \ce{^{13}C}, Al, Cu, and Au targets. The beam energy is 908\,keV. The parameters from the linear fit are used for the calibration of the spectra measured in the irradiation setup.}
\label{fig:cal}
\end{figure*}

The only reasonable possibility for a calibration is the use of backscattered protons from known targets. For this purpose, the SSBs have been mounted in the D-shaped chamber at beam line 2 that has been used for the RBS measurements (cf. Figure~\ref{fig:setup_picture}). The electronics configuration including all cabling remains the same as in the irradiation setup. The only difference is the use of a different vacuum feed-through, which could introduce a slight change in the capacitance between detector and preamplifier. One by one the SSBs have been calibrated with four different known targets. \ce{^{13}C}, aluminum, copper, and gold targets are easily available and cover a wide energy range of backscattered protons without the need for changing the beam energy (cf. Tab.~\ref{tab:cal_E}). To determine the ADC channel-to-energy mapping, the positions of the high energy edges of the spectra are used. An example of the calibration is presented in Figure~\ref{fig:cal}.

\begin{table}[tb]
\centering
\caption{Values for the high energy edge of $165^\circ$ backscattered protons from different target materials for 908\,keV beam energy. Without changing the beam energy, a range of more than 200\,keV is covered.}
\label{tab:cal_E}
\begin{tabular}[c]{cc}\\
		\toprule
		Target material & Proton energy \\
		 & (keV) \\ \midrule
		\ce{^{13}C} & 669.2 \\
		Al & 784.0 \\
		Cu & 854.3 \\
		Au & 889.9 \\
		\bottomrule
\end{tabular}
\end{table}

A consistency check has been performed by measuring the proton spectrum in the irradiation setup with four detectors that have been calibrated with this method. The largest energy offset between two detectors is less than 12\,keV. This indicates an accuracy of the calibration that is more than suitable for the measurement of energy spectra for irradiation purposes.

% --------------------------------------------------------------------------------------

\section{Irradiation of SDDs for \textit{LOFT}}\label{sec:LOFT}
The described setup has been used so far for two irradiation campaigns, in which prototypes of the detectors that will be used on-board \textit{LOFT} have been tested for radiation hardness. In the following, a brief introduction of the \textit{LOFT} project is given and the approach for the irradiation campaigns is presented.

\subsection{\textit{LOFT} mission overview}
\textit{LOFT} is a newly proposed space mission, which was recently selected by ESA as one of five candidates for the M3 mission of the Cosmic Vision program. These five mission concepts compete for a launch opportunity in the early 2020s. The scientific focus of \textit{LOFT} is to answer fundamental questions about the motion of matter orbiting close to the event horizon of a black hole and the state of matter in neutron stars \cite{LOFT}.

The satellite will operate in a near-equatorial low Earth orbit ($\sim 600\,\mathrm{km}$ altitude, $<5^\circ$ inclination). It will be equipped with two scientific instruments: the \textit{Large Area Detector} (\textit{LAD}) \cite{LAD} and the \textit{Wide Field Monitor} (\textit{WFM}) \cite{WFM}. The \textit{LAD} covers a geometric area of $\sim18\,\mathrm{m^2}$ that leads to an unprecedented effective area of $\sim10\,\mathrm{m^2}$ at 8\,keV X-ray energy. The narrow field-of-view (FOV) of $<1^{\circ}$ is defined by novel microcapillary plate X-ray collimators. Complementary to the \textit{LAD} is the \textit{WFM}, a coded mask instrument with a large FOV that covers about 1/3 of the sky. Its main purpose is to provide sources to point at with the \textit{LAD} and to catch transient and bursting events. 

Both instruments use the same solid-state detector, a slightly modified version of the Silicon Drift Detector (SDD), which was originally developed for \textit{ALICE} at the \textit{LHC} at CERN and is now implemented in its inner tracking system \cite{ALICE_SDD}. This detector offers a small mass-per-area ratio ($\sim1\,\mathrm{kg m^{-2}}$) and an energy resolution of 260\,eV at 6\,keV. For X-ray detection the thickness of the sensitive layer has been increased from 300\,\textmu m to 450\,\textmu m. The main difference between the \textit{LAD} and the \textit{WFM} SDDs is the anode pitch ($\sim 1\,\mathrm{mm}$ for the \textit{LAD}, $\sim 0.15\,\mathrm{mm}$ for the \textit{WFM}).

\subsection{Irradiation campaigns}
Up to now, two irradiation campaigns for \textit{LOFT} SDD prototypes have been carried out with the setup in Tübingen, the first in June 2012, and the second in December 2012. The goal was to determine the degradation of the energy resolution by soft protons during the proposed mission time of five years. Further irradiations with more energetic protons ($\sim 50\,\mathrm{MeV}$) have been carried out at other facilities.

\begin{figure*}[tb]
\centering
\includegraphics[width=.5\textwidth]{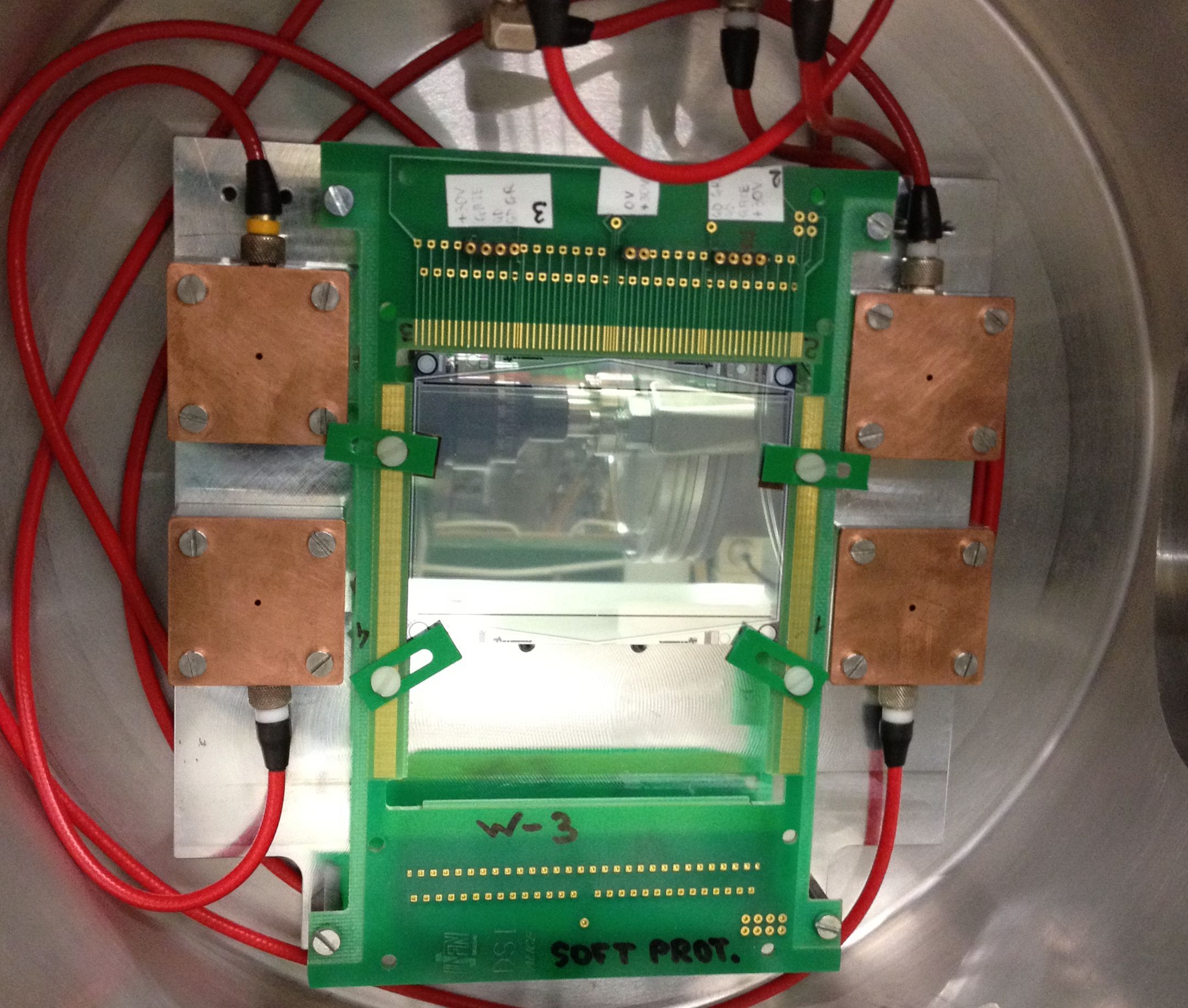}
\caption{Picture of a \textit{LOFT} detector prototype on a printed circuit board (PCB), mounted in the irradiation chamber with four monitor detectors around. The connectors on the upper part of the PCB have been used to bias the SDD and test structures around the SDD during the irradiation.}
\label{fig:LOFT_PIC}
\end{figure*}

The irradiated SDDs have half the size of the final version, and a different anode pitch on each side. A picture of one of the detector prototypes is shown in Figure~\ref{fig:LOFT_PIC}. During irradiation, a bias voltage of 30\,V has been applied. The fluence calculation takes into account that the LAD and the WFM have different solid angles, and different materials (optical filters, debris shields, etc.) are placed in front of the detectors.

Since readout electronics are not available yet, the detector characterization is limited to a measurement of the leakage current before and after irradiation. Nevertheless, the combination of the leakage current measurement and additional information, which is gained from the measurement of test structures (gated diodes and MOS capacitors) that are placed in the detector corners, yields a well founded statement on the radiation hardness. A publication that presents the results of the soft proton irradiation campaigns together with the results of other irradiations is in preparation. In the following sections \ref{sect:1st_camp} and \ref{sect:2nd_camp} the irradiation procedures are described, and the nominal and the applied fluences are summarized.

\subsubsection{First preliminary irradiation campaign}\label{sect:1st_camp}
In the first campaign two proton energies were used: $300\pm33\,\mathrm{keV}$ to maximize ionization in the insulating field oxide on the SDD surface, and $\sim 838\pm52\,\mathrm{keV}$, because it is the maximal energy available for a uniform flux distribution. The fluence has been applied in four steps with intermediate measurements of the I-V-curves of the gated diodes, and leakage current measurements before and after the entire irradiation. The fluence for each step is selected to reach a defined NIEL for the WFM and the LAD: in Step 1 half the NIEL expected for the LAD is reached, Step 2 increases the NIEL to the orbital value for the LAD, Step 3 reaches approximately the orbital NIEL for the WFM, and finally Step 4 increases the NIEL to twice the orbital value for the WFM. The irradiations were performed on three sequent days and the irradiation durations for the individual steps ranged from several seconds to more than ten minutes. The nominal fluences have been calculated for slightly lower energies ($208\,\mathrm{keV}$ and $820\,\mathrm{keV}$) than the ones applied. The nominal and the applied fluence of each step are listed in Table~\ref{tab:1st_camp}. 

\begin{table*}[tb]
\centering
\caption{Nominal fluences $\Phi_\mathrm{nom}$ and applied fluences $\Phi_\mathrm{app}$ of the four irradiation steps of the first irradiation campaign. The nominal fluences are calculated for 208\,keV and 820\,keV, while during the irradiation 300\,keV and 838\,keV have been used. The total nominal dose corresponds to 2.0 times the orbital NIEL for the WFM, and to 14.9 times the orbital NIEL for the LAD.}
\label{tab:1st_camp}
\begin{tabular}[c]{ccccc}
\\\toprule
Irradiation	&	$\Phi_\mathrm{nom, 208\,keV}$	&	$\Phi_\mathrm{app, 300\,keV}$	&	$\Phi_\mathrm{nom, 820\,keV}$	&	$\Phi_\mathrm{app, 838\,keV}$	\\
step				&	($\mathrm{cm^{-2}}$)					&	($\mathrm{cm^{-2}}$)					&	($\mathrm{cm^{-2}}$)					&	($\mathrm{cm^{-2}}$)					\\
\midrule
1						&	$3.50 \cdot 10^6$						&	$3.38 \cdot 10^6$						&	$7.85 \cdot 10^5$						&	$9.53 \cdot 10^5$						\\
2						&	$3.50 \cdot 10^6$						&	$3.44 \cdot 10^6$						&	$7.85 \cdot 10^5$						&	$6.51 \cdot 10^5$						\\
3						&	- - -													&	- - - 												&	$2.40 \cdot 10^7$						&	$2.24 \cdot 10^7$						\\
4						&	$1.10 \cdot 10^5$						& $1.36 \cdot 10^5$						&	$4.28 \cdot 10^7$						&	$4.34 \cdot 10^7$						\\
\midrule
total				&	$7.11 \cdot 10^6$						&	$6.96 \cdot 10^6$						&	$6.84 \cdot 10^7$						&	$6.74 \cdot 10^7$ 						\\
\bottomrule
\end{tabular}
\end{table*}

\subsubsection{Second irradiation campaign}\label{sect:2nd_camp}
For the second campaign two SDD prototypes from different development stages have been irradiated. Prototype 1 is the same detector that has been used in the first campaign. Meanwhile, the induced radiation damages have annealed. The second SDD is the latest prototype for \textit{LOFT}. The irradiation procedure was simplified compared to the first campaign by using just one proton energy ($838\pm53\,\mathrm{keV}$) and applying the total fluence in only one step. Since the fluence was calculated to induce ten times the orbital WFM NIEL, the flux has been increased as much as possible to minimize the duration, and, therefore, avoid annealing effects during the irradiation. Both detector prototypes have been irradiated on the same day and each irradiation took less than half an hour. The nominal and applied fluences for the two detectors are given in Table~\ref{tab:2nd_camp}.

\begin{table*}[tb]
\centering
\caption{Nominal fluences $\Phi_\mathrm{nom}$ and applied fluences $\Phi_\mathrm{app}$ for the two SDD prototypes used in the second irradiation campaign (Prototype 1 is the same detector that has been used in the first campaign). The nominal fluences correspond to 10 times the orbital NIEL expected for the WFM (74.5 times LAD).}
\label{tab:2nd_camp}
\begin{tabular}[c]{ccc}
\\\toprule
SDD					&	$\Phi_\mathrm{nom, 838\,keV}$	&	$\Phi_\mathrm{app, 838\,keV}$	\\
prototype		&	($\mathrm{cm^{-2}}$)					&	($\mathrm{cm^{-2}}$)					\\
\midrule
1						&	$3.59 \cdot 10^8$						&	$3.73 \cdot 10^8$						\\
2						&	$3.59 \cdot 10^8$						&	$3.62 \cdot 10^8$						\\
\bottomrule
\end{tabular}
\end{table*}

% --------------------------------------------------------------------------------------

\section{Conclusions}\label{sec:Concl}
A setup for the irradiation of solid-state detectors with soft protons has been constructed at the accelerator facility of the University of Tübingen. It has already proven its applicability for radiation hardness tests of X-ray detectors for future space missions.

The monitoring system with silicon surface barrier detectors enables the test of new detector prototypes under orbital radiation conditions, even in early development phases when readout electronics are not yet available. The applied methods for the energy calibration and the determination of the effective areas of the monitor detectors show consistent results. The contribution of small angle scattered protons is effectively reduced with additional apertures. Another key advantage of the setup is the high beam uniformity with a flux deviation of less than 3\% over an area of $8 \times 8\,\mathrm{cm^2}$. Therefore, the setup is especially suitable for the irradiation of large detectors up to diameters of about 14\,cm.

Two irradiation campaigns for the \textit{LOFT} project have been carried out so far, in which the desired fluences have been reached within a few percent. A publication that presents the results of these campaigns together with the results of other irradiations of \textit{LOFT} detector prototypes is in preparation. Further applications of the setup for \textit{LOFT} detector prototypes are likely, and the setup is available for the irradiation of other solid-state detectors as well, e.g. the \textit{SVOM}\footnote{Space-based multi-band astronomical Variable Objects Monitor} CCDs or the CdZnTe detectors that are intended for the use on-board \textit{MIRAX}\footnote{Monitor e Imageador de RAios-X}.

Underway is a modification and extension of the irradiation setup that will give the possibility to measure small angle reflection rates of soft protons under grazing incidence. The scattering targets will be X-ray mirror shells, like the ones used on \textit{eROSITA}\footnote{extended ROentgen Survey with an Imaging Telescope Array}. The goal is to achieve an energy and angular resolution that is sufficient to constrain the underlying physical process. The results will be beneficial for simulations of soft proton background and detector damage on X-ray observatories with focusing Wolter-type optics.

% --------------------------------------------------------------------------------------

\section*{Acknowledgment}

This work is partially supported by the Bundesministerium für Wirtschaft und Technologie through the Deutsches Zentrum für Luft- und Raumfahrt (Grant FKZ 50 OO 1110).

Thanks to the Universität Tübingen for the support of the Kepler graduate school.

\textit{LOFT} is a project funded in Italy by ASI under contract ASI/INAF n. I/021/12/0.

The development of SDDs for X-ray detection is going on under INFN R\&D projects. \textit{LOFT} is the first application of these devices.



%% References


%\bibliographystyle{elsarticle-num}






% \end{thebibliography}

\end{document}